\documentclass[%
 reprint,
 amsmath,amssymb,
 aps,
]{revtex4-2}

\usepackage{graphicx}
\usepackage{dcolumn}
\usepackage{bm}
\usepackage{xcolor}

\begin{document}

\preprint{APS/123-QED}

\title{Scaling Breakdown as a Signature of Spinon-Gauge Interaction in the Quantum Spin Liquid  YbZn$_2$GaO$_5$}

\author{
Shannon Gould$^{1}$,
John Singleton$^{2}$,
Rabindranath Bag$^{3}$,
Sara Haravifard$^{3,4,5}$,
Sheng Ran$^{1}$ \\
\textit{{\small $^1$Department of Physics, Washington University in St. Louis, St. Louis, Missouri 63130, USA} \\
{\small $^2$National High Magnetic Field Laboratory, Los Alamos National Laboratory, Los Alamos, New Mexico 87545, USA} \\
{\small $^3$Department of Physics, Duke University, Durham, North Carolina 27708, USA} \\
{\small $^4$Department of Mechanical Engineering and Materials Science, Duke University, Durham, North Carolina 27708,~USA}
{\small $^5$Department of Electrical and Computer Engineering, Duke University, Durham, North Carolina 27708, USA}
}}

\date{\today}

\begin{abstract}
Scaling behavior in magnetization has been reported in a wide range of quantum spin liquid (QSL) candidates and is often interpreted as evidence for scale-free spin liquid physics. Here we present a comprehensive scaling analysis of high-field magnetization measurements on the QSL material YbZn$_2$GaO$_5$. Between 5 K and 70 K, $M(H)$ displays scale invariance resembling that of a zero-field quantum critical point. Below 3 K, we observe a breakdown of this scale invariance that cannot be recovered by simply changing the critical exponents. This temperature coincides with the onset of enhanced spin correlations observed in $\mu$SR measurements. Moreover, the form of the deviation from scaling is consistent with collective spinon excitations coupled via emergent gauge interactions. These results indicate that the breakdown of scaling reflects the emergence of intrinsic low-energy excitations upon entering the QSL regime. Our work clarifies that magnetic scaling is associated with quantum critical fluctuations rather than with the spin liquid phase itself, and establishes magnetization scaling as a sensitive thermodynamic probe of emergent energy scales in QSL systems.

\end{abstract}

\maketitle

Quantum spin liquids (QSLs) are magnetically disordered states characterized by long-range quantum entanglement and fractionalized spin excitations \cite{Balents2010,Savary2016,Knolle2019,Wen2019,Broholm2020,Clark2021,Zhou2017}. Unlike conventional magnets, QSLs do not break spin rotational symmetry and lack a conventional ordering temperature \cite{Balents2010,Savary2016,Knolle2019,Wen2019,Broholm2020,Clark2021}. The absence of long-range order suggests that such systems may not possess an intrinsic symmetry-breaking energy scale, raising the question of whether thermodynamic responses in a QSL should exhibit scale-invariant behavior \cite{Balents2010,Broholm2020,Clark2021,Shaginyan2010,Shaginyan2020}.

Indeed, scale invariance in magnetization and susceptibility has been reported in several QSL candidates \cite{Isono2016,Helton2010,Khatua2022} and is frequently interpreted in terms of quantum critical (QC) scaling. In such a scenario, the system is governed by a single characteristic energy scale, typically expressed through the dimensionless ratio $g\mu_BH/k_BT$, and thermodynamic quantities collapse onto universal scaling functions \cite{Vojta2003,Lohneysen2007,Sachdev2011,Sachdev_QuantumPhaseTransitions,Paschen2021}. This behavior is often associated with proximity to a zero-field quantum critical point (QCP) and has been considered compatible with, or even indicative of, QSL physics.

However, it remains unclear whether scale invariance should persist inside a genuine QSL regime. If fractionalized spinon excitations are coupled to emergent gauge fields, their intrinsic many-body dynamics may introduce an additional low-energy scale unrelated to symmetry breaking. In that case, the QSL phase itself would not be strictly scale-free, and single-parameter quantum critical scaling could break down upon entering the spin liquid regime. Despite extensive studies of scaling behavior in QSL candidates \cite{Isono2016,Modic2021,Tokiwa2013,Tokiwa2014,Helton2010,Majumder2020,Murayama2022,Kimchi2018_ScalingAnd,Shaginyan2013,Hong2021,Peng2025,Liu2018,Kundu2020,Song2021,Han2023}, a systematic breakdown of such scaling attributable to intrinsic QSL excitations has not been experimentally demonstrated. 

In this work, we report high-field magnetization measurements of the QSL candidate YbZn$_2$GaO$_5$ in pulsed magnetic fields of up to 60 T and over a temperature range from 0.6 K to 160 K. Between 5 K and 70 K, the magnetization exhibits robust $H/T$ scaling consistent with zero-field quantum critical behavior. Upon cooling below approximately 3 K, this scale invariance breaks down and cannot be restored by modifying critical exponents or redefining scaling variables. This temperature scale coincides with the onset of enhanced spin correlations observed in $\mu$SR \cite{Wu2025} and matches the low-energy excitation scale identified by inelastic neutron scattering (INS) \cite{Bag2024}, indicating that the scaling breakdown reflects the emergence of intrinsic low-energy excitations within the QSL regime. We further show that the observed deviations are captured by a phenomenological description incorporating an emergent energy scale associated with collective spinon dynamics.

Our results provide thermodynamic evidence for spinon–gauge dynamics in a quantum spin liquid. They clarify that the observed magnetic scaling is associated with quantum critical fluctuations rather than with the spin liquid phase itself, and demonstrate that high-field magnetization scaling serves as a powerful probe of emergent energy scales and fractionalized excitations in QSL materials.

YbZn$_2$GaO$_5$ is related to the Yb(Mg,Zn)GaO$_4$ triangular lattice compounds that have been well-studied as possible QSL candidates \cite{Xu2016,Shen2016,Paddison2017,Li2017SpinonFermi, Li2017NearestNeighbor, Shen2018, Li2019, Steinhardt2021Constraining, Ma2018, Steinhardt2021PhaseDiagram}. YbZn$_2$GaO$_5$ likewise is a QSL candidate exhibiting no long-range magnetic ordering down to the lowest measurable temperatures \cite{Bag2024,Wu2025,Zhao2025}. A broad, continuum-like INS spectrum was observed that possesses a  gap at the $\Gamma$ point but appears to remain gapless between the $M$ and $K$ points. This, along with the $T^2$-dependence of the heat capacity, suggests a U(1) Dirac QSL state below 0.15 K. Theoretical comparison with the experimental INS spectra \cite{Bag2024} indicates that a $J_1-J_2$ XXZ model explains the behavior well, with the spinon energy scale depending on the dimensionless parameter $J_2/J_1$. $\mu$SR measurements further confirmed the absence of long-range magnetic order and revealed enhanced spin correlations below 3~K~\cite{Wu2025}.

Our magnetization measurements were performed down to 0.6 K, which remains above the temperature scale proposed for the U(1) Dirac QSL regime \cite{Bag2024}, yet well below 3 K—the onset of enhanced spin correlations observed in $\mu$SR measurements \cite{Wu2025}. Magnetization for both field directions and for temperatures down to 0.6 K is shown in Fig. \ref{fig:MvsH}. At the base temperature, for both directions, the magnetization increases rapidly at low fields, bends over around 10 T, and then approaches saturation. A small residual increase in magnetization at high fields arises from the van Vleck contribution associated with excited CEF states, which is linear in field (SI Fig. 1).

\begin{figure}
\centering
\includegraphics[width=0.48\textwidth]{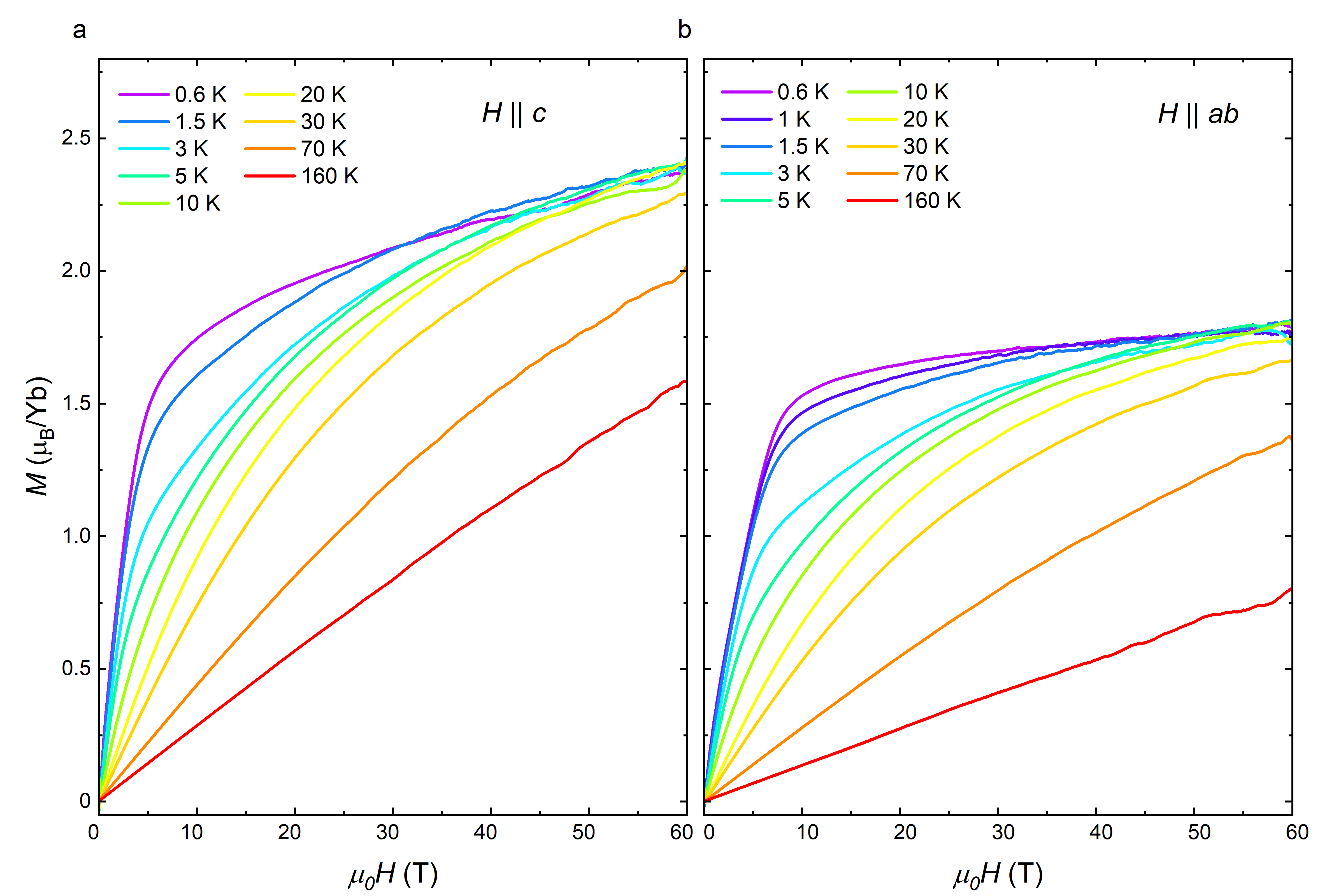}
\caption{Magnetization as a function of magnetic field for (a) $H||c$ and (b) $H||ab$.}
\label{fig:MvsH}
\end{figure}

As temperature increases, a noticeable change occurs around 3 K: below this temperature, the magnetization curve bends over just before 10 T, whereas above 3 K, no clear bending is observed. This change, which is associated with anisotropy in the $g$ factors, is more clearly reflected in the magnetic susceptibility, $\chi(H)$, calculated as $\partial M/\partial H$ (SI Fig. 2). Below 3 K, $\chi(H)$ exhibits a distinct shoulder feature before 10 T, which disappears at higher temperatures.

To further investigate the evolution of magnetization, we performed a scaling analysis after subtracting the van Vleck contribution. Figure \ref{fig:Scaled Magnetization_Inset_StDev Plots} shows plots of $(M-M_{vv})T^{\gamma-1/\nu z}$ versus $\mu_0(H-H_c)/T^{1/\nu z}$, where $M_{vv}=\chi_{vv}(\mu_0 H)$. We find that for particular choices of $\gamma,\nu z,H_c$ and for both field directions, all data points between 5 K and 70 K collapse onto a single universal curve spanning more than two decades along the $y$-axis and nearly four decades along the $x$-axis. Above 70 K, e.g., at 160 K, the magnetization fits well to a Brillouin function (SI Fig. 3), indicating the system is in the non-interacting paramagnetic regime.

\begin{figure*}
\centering
\includegraphics[width=1\textwidth]{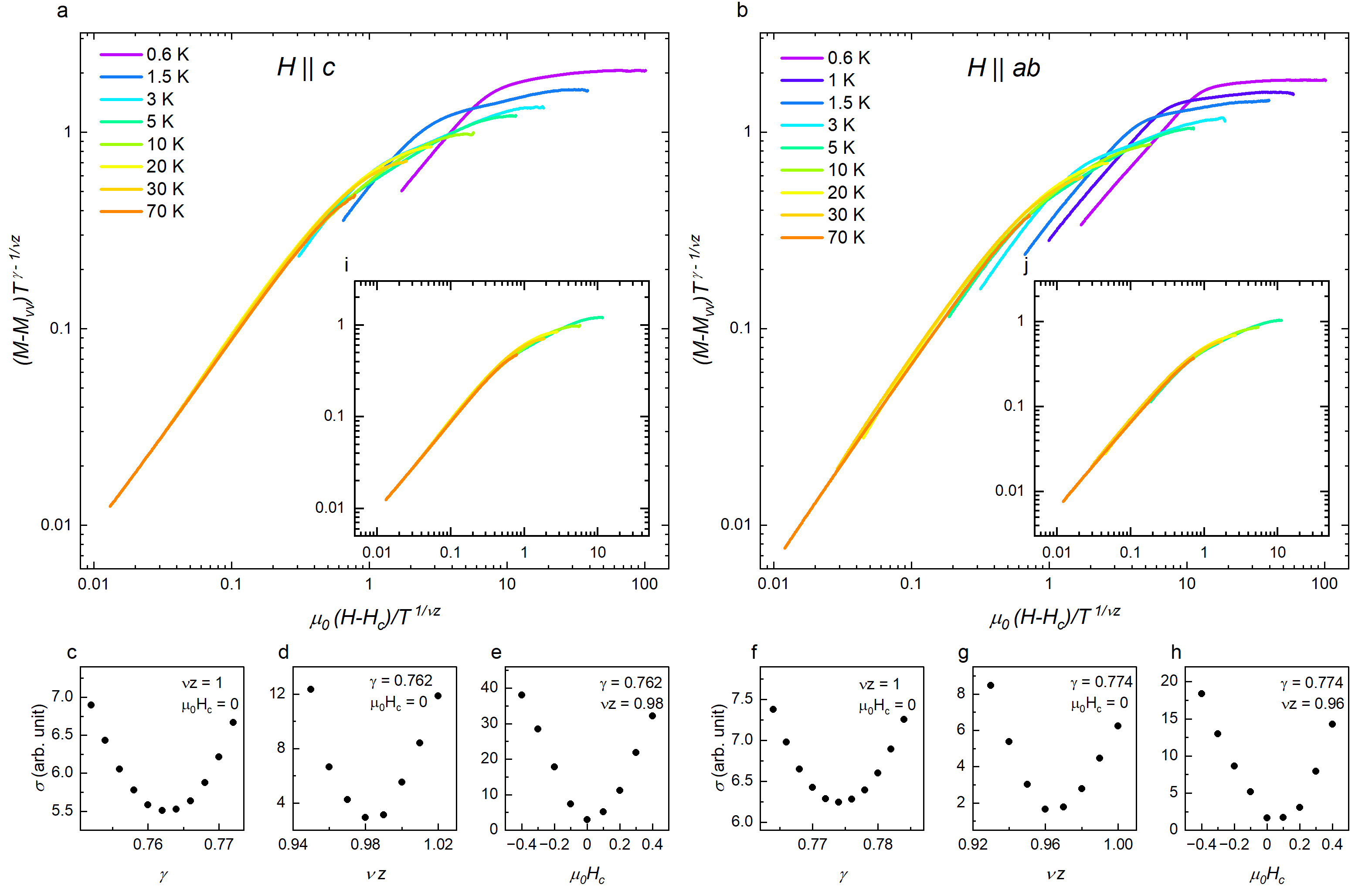}
\caption{Magnetization scaled according to $(M-M_{vv})T^{\gamma-1/\nu z}=f[\mu_0(H-H_c)/T^{1/\nu z}]$, with $M_{vv}=\chi_{vv}(\mu_0 H)$, for (a) $H||c$ and (b) $H||ab$. All temperatures up to 70 K are shown in (a) and (b). Insets (i) and (j) show only the temperatures that collapse onto a universal curve. The optimization procedure for each critical exponent involved adjusting its value in order to minimize the standard deviation, $\sigma$, across the curves. Details are provided in the SI, and results of the optimization procedure are shown in (c)-(e) for $H||c$ and (f)-(h) for $H||ab$.}
\label{fig:Scaled Magnetization_Inset_StDev Plots}
\end{figure*}

Scale invariance in magnetization and other thermodynamic quantities has been reported in several other quantum spin liquids, attributed to different mechanisms including quantum criticality, quenched disorder effects, and local frustration of the exchange interactions \cite{Isono2016,Modic2021,Tokiwa2014,Tokiwa2013,Helton2010,Murayama2022,Shaginyan2013,Khatua2022,Kundu2020,Song2021}. Although we cannot exclude other mechanisms, our results resemble those reported for quantum criticality. For this scale invariance observed in $M$ to be explained as QC scaling, we expect the same critical exponent values for the two field orientations. In our case, the values of $\gamma$, 0.762 for $H||c$ and 0.774 for $H||ab$, differ by less than 2\%. A much larger difference in critical exponents would be expected for a system where disorder dominates \cite{Vojta2003,Isono2016}. 

The extracted exponent $\nu z=0.98$ ($0.96$) for $H||c$ ($H||ab$) agrees well with the value $\nu z=1$ predicted by the itinerant Hertz–Millis–Moriya theory for antiferromagnetic quantum criticality \cite{Zhu2003,Tokiwa2014}. The exponent $\gamma = 0.76$ ($0.77$) is also comparable to values reported in other quantum spin liquids exhibiting scaling behavior due to quantum criticality, such as the organic spin liquid $\kappa$-(BEDT-TTF)$_2$Cu$_2$(CN)$_3$~\cite{Isono2016}.

In addition, like $\kappa$-(BEDT-TTF)$_2$Cu$_2$(CN)$_3$~\cite{Isono2016}, the scaled magnetization reveals two distinct regimes as a function of $H/T$. In the low $H/T$ regime, $(M-M_{vv})T^{\gamma-1/\nu z}$ increases linearly with $H/T$, whereas in the high $H/T$ regime, $(M-M_{vv})T^{\gamma-1/\nu z}$ follows a sublinear power law dependence. This behavior is more clearly seen in the scaled susceptibility (SI Fig. 4). $(\chi - \chi_{vv}) T^{\gamma}$ remains constant at low $H/T$, while at high $H/T$, the universal curve follows $(\chi - \chi_{vv}) T^{\gamma} \approx (H/T)^{-0.8}$. The crossover between these two regions has been attributed to a boundary of the QC region in $H-T$ phase space~\cite{Isono2016}. Near a QCP, a fan-shaped region extends from the zero-temperature critical point where physical properties are governed by thermal excitations of the quantum critical ground state \cite{Vojta2003,Sachdev_QuantumPhaseTransitions,Sachdev2011,Paschen2021}. The boundaries of this region arise from the competition between thermal energy and the energy scale of critical fluctuations, beyond which the system crosses over into a quantum disordered regime \cite{Vojta2003}. For a zero-field QCP, only one side of the fan is accessible in the $H-T$ phase diagram, so the observed crossover in SI Fig. 4 is consistent with moving from the QC region into the quantum disordered region.

To extract the locations of this boundary, we performed independent linear fits to these two regimes and interpreted the location of their intersection as the boundary point for each temperature.  The crossover points for $H||c$ are shown in Fig. \ref{fig:HparaC Contour Zoom}, which plots the $y$-value of Fig. \ref{fig:Scaled Magnetization_Inset_StDev Plots}a as the color contour. For $T \geq 5$ K, the crossover points trace back to the origin at $0$ K and $0$ T (black dashed line in Fig. \ref{fig:HparaC Contour Zoom}).

\begin{figure}
\centering
\includegraphics[width=0.48\textwidth]{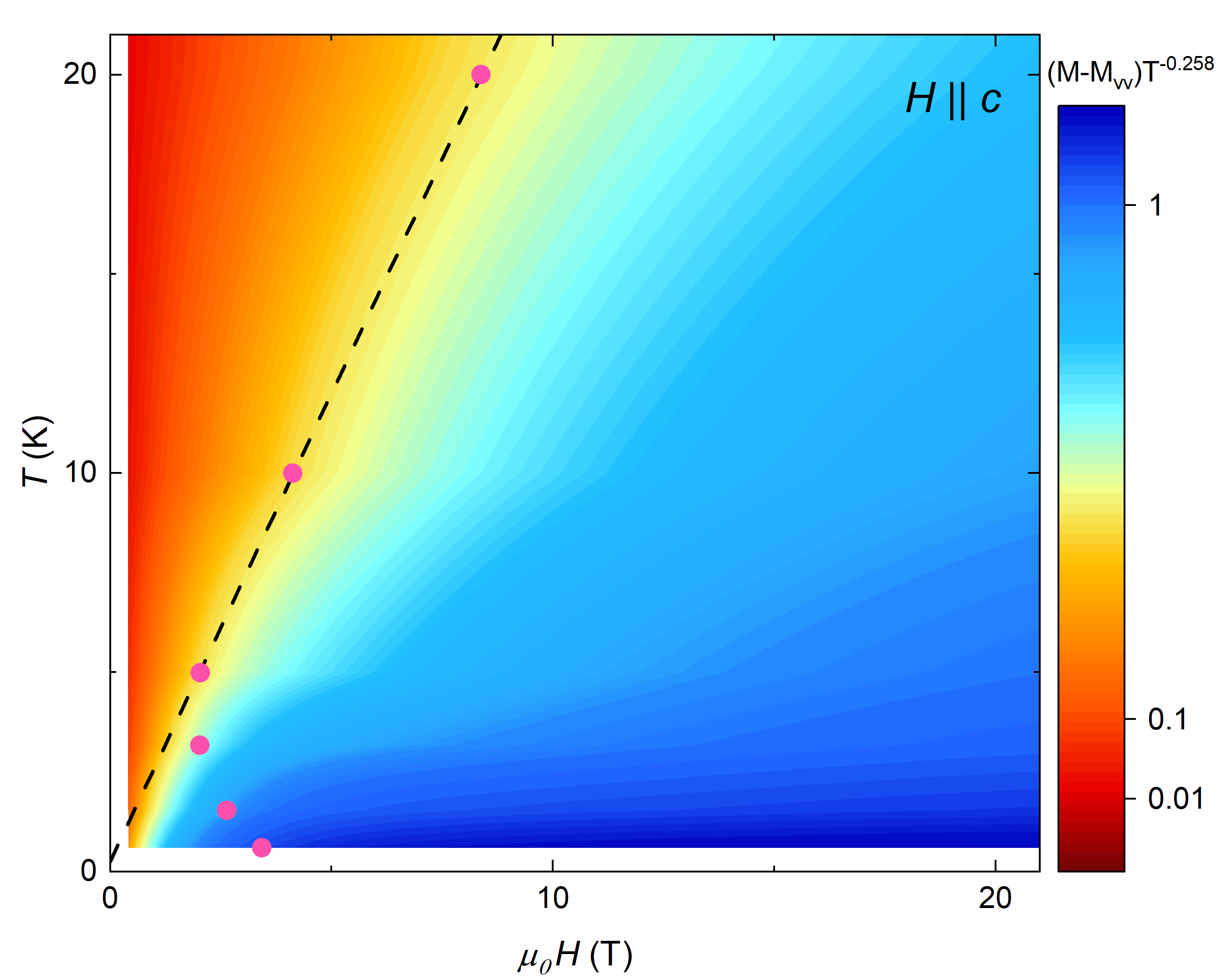}
\caption{Phase diagram for $H||c$, zoomed in to low $H-T$ region. The scaled magnetization, given by $(M-M_{vv})T^{\gamma - 1/\nu z}$ is plotted as the color. Scatter points on the plot indicate the crossover where $H/T$ dependence changes.}
\label{fig:HparaC Contour Zoom}
\end{figure}

While scaling behavior has been widely observed in quantum spin liquid materials, the most striking feature revealed by our scaling analysis (Fig. \ref{fig:Scaled Magnetization_Inset_StDev Plots}) is the breakdown of scale invariance beginning around 3 K. For both field orientations, the low temperature data deviate from the universal curve systematically. At 3 K, the deviation is small but most apparent in the small-x and large-x limits of the curve. As temperature is lowered, the curves progressively shift upwards and to the right, relative to the universal curve. This breakdown of the scaling is also evident in Fig. 3 where the crossover points deviate significantly from the high temperature line below 5~K.

The scaling behavior above 3~K indicates that the system is governed by a single characteristic energy scale. In the vicinity of a QCP, the correlation length and correlation time of the order parameter fluctuations diverge as $\xi \propto |H-H_c|^{-v}$ and $\xi_\tau \propto \xi^z$, respectively, giving rise to a single characteristic energy scale $E^* \sim \xi_\tau^{-1} \propto |H-H_c|^{v z}$ and scale invariance in certain thermodynamic quantities \cite{Vojta2003,Lohneysen2007,Sachdev2011}. The breakdown of scaling below 3~K therefore signals the emergence of an additional low-energy scale, beyond that associated with quantum critical fluctuations.

A trivial origin of an additional energy scale would be the development of magnetic order, which has indeed been shown to destroy scaling behavior in certain quantum spin liquid candidates \cite{Isono2016, Modic2021, Tokiwa2014}. In YbZn$_2$GaO$_5$, however, no long-range order has been detected by thermodynamic measurements, neutron scattering, or $\mu$SR down to the lowest measured temperatures \cite{Bag2024, Zhao2025, Wu2025}. Instead, the temperature scale of $\approx3$~K coincides with the onset of enhanced spin correlations reported by $\mu$SR \cite{Wu2025}. This temperature is also consistent with the excitation energy of $\approx0.3$~meV observed in inelastic neutron scattering~\cite{Bag2024}. The correspondence between these independent energy scales strongly suggests that the breakdown of scaling is not driven by conventional ordering, but rather reflects the emergence of intrinsic low-energy excitations associated with the quantum spin liquid state.

In principle, an additional energy scale can arise intrinsically within certain classes of quantum spin liquids, particularly those supporting gapless emergent gauge fluctuations \cite{Polchinski1994,Motrunich2005,Lee2006,Wen2002,Senthil2004,Shaginyan2020}. In such systems, the coupling between fractionalized excitations and gauge fields may introduce a new low-energy scale beyond that associated with quantum critical scaling, thereby leading to a breakdown of scale invariance. However, to the best of our knowledge, a breakdown of scaling behavior attributable to intrinsic low-energy excitations of a quantum spin liquid has not been previously reported.

\begin{figure}
\centering
\includegraphics[width=0.48\textwidth]{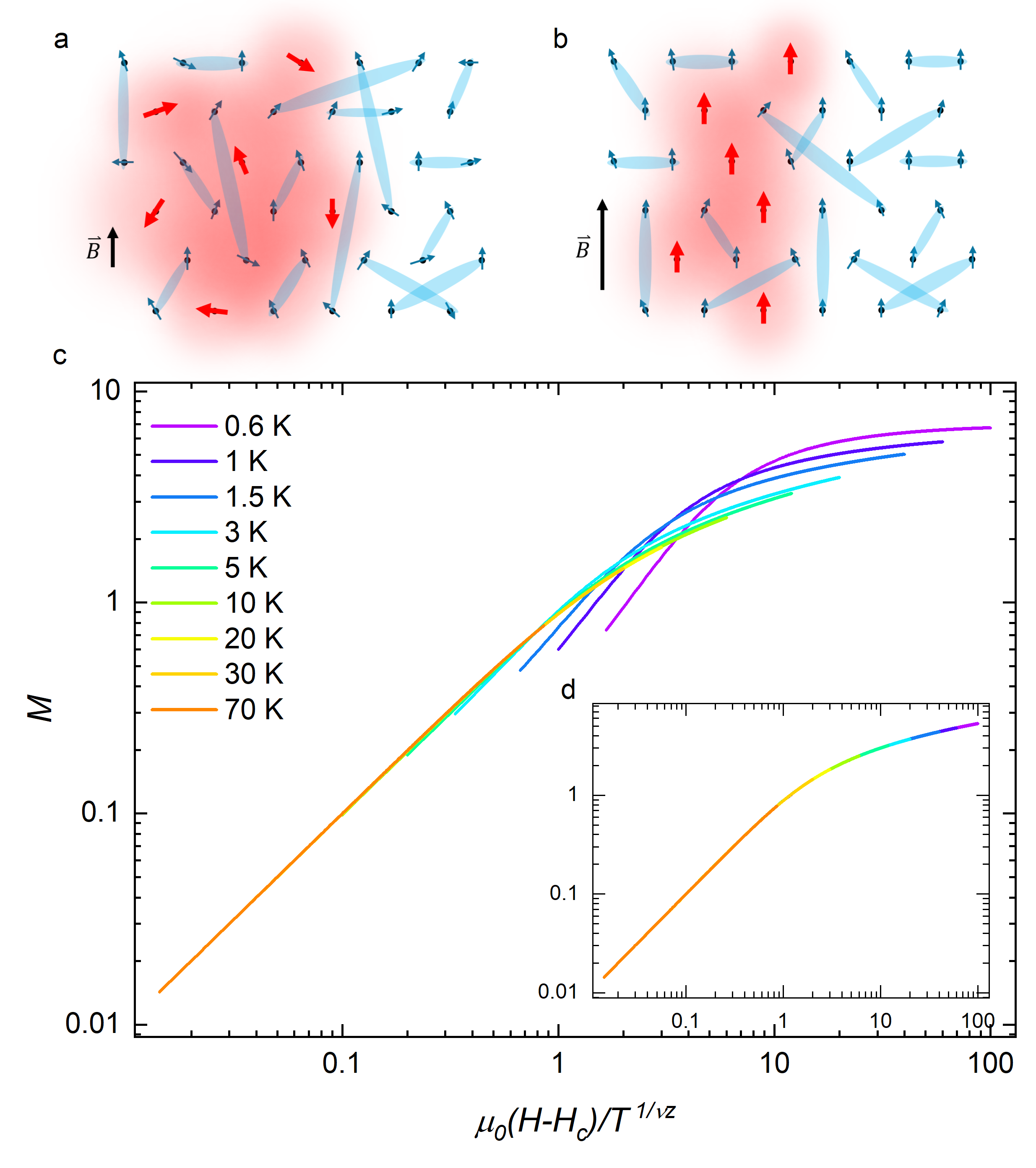}
\caption{
(a) Cartoon diagram of a QSL state with spinon excitations under a weak magnetic field. Spins (blue arrow) reside on a triangular lattice and exhibit long-range entanglement via singlet bonds (light blue bubble). Broken singlet bonds result in a spinon excitation (red arrow). Under weak magnetic field, spins partially align with the field. The spinon excitations, due to their collective nature, cannot be polarized in such a weak field.
(b) Under a stronger magnetic field, the spinon excitations are collectively polarized. This enhances the magnetization of the state beyond that expected of a system without spinon excitations.
(c) Model for magnetization with the inclusion of an emergent low-energy scale.
(d) Model for magnetization within the quantum critical regime, scaled using $\gamma=0.8$, $\nu z=1$, $H_c=0$.
}
\label{fig:Scaling Model}
\end{figure}

The breakdown of scaling arising from low-energy excitations of a quantum spin liquid can be understood intuitively in terms of spinon excitations coupled to an emergent internal gauge field. The interactions from internal gauge field are very nontrivial, and they result in a continuum of states possessing an entangled, many-body nature \cite{Savary2016,Broholm2020,Motrunich2005,Polchinski1994}. When subject to an external magnetic field, this results in a departure from the behavior of a system with magnon excitations. A small magnetic field is insufficient to polarize the spinon quasiparticles independently because they form a collective, entangled state. However, at sufficiently large magnetic fields, the spinon excitation continuum is redistributed, leading to polarization of the collective spinon excitations and a magnetization exceeding that expected for a system without spinon quasiparticles.

Examining our data, we find that they exhibit all the key features expected for a scaling breakdown driven by intrinsic low-energy excitations of a quantum spin liquid \cite{Senthil2004,Han2023,Shen2016,Yamashita2010}: 1. the breakdown occurs below the characteristic energy scale associated with the spin liquid excitations; 2. the magnetization is suppressed at small $H/T$ compared to the high-temperature scaling form; 3. the magnetization becomes enhanced at large $H/T$ compared to the high-temperature scaling form.

To describe the breakdown of scale invariance in a more quantitative manner, we introduce a phenomenological model for $M$ based on this intuitive physics picture.  In the quantum critical regime, $M$ only has the critical contribution: 
\[
M=M_{QC}=T^{-\gamma+1/\nu z}\Phi[\mu_0(H-H_c)/T^{1/\nu z}],
\]
where $\Phi(x)$ is an arbitrary scaling function. This model inherently captures the lack of any relevant energy scale in the system. We choose the scaling function to be the hypergeometric function (see SI for details). We plot the scaled quantities, $MT^{\gamma-1/\nu z}$ versus $\mu_0(H-H_c)/T^{1/\nu z}$ in Fig. \ref{fig:Scaling Model}d. Representative values for the critical exponents, $\gamma=0.8$ and $\nu z=1, H_c=0$, are used. As expected, this model results in complete collapse of all temperatures over the full x-range.

Next, we incorporate an emergent energy scale, $\Delta$, into the model that competes with the scale invariance arising from quantum critical fluctuations. $M$ is modeled as 
\[
M=M_{QC}[1+B(T)f(X)].
\]
Two new functions, $B(T)$ and $f(X)$, encode the contributions from low-energy excitations associated with the spin liquid state. $B(T)$ controls the temperature dependence of this additional contribution. The function $f(X)$ captures the field dependence of the spinon contribution, which is negative at low magnetic field and positive at high magnetic field (See SI for details). The functional forms of $B(T)$ and $f(X)$ allow some flexibility in selection of the parameters so that the details of the low temperature curves can be adjusted. However, the key features of the functions, including their limits, are fixed regardless of specific parameter values.

In Fig. \ref{fig:Scaling Model}c, we show this model applied to $M$ for all temperatures studied in the experiment. The general behavior of the low temperature curves agrees qualitatively with the data shown in Fig. \ref{fig:Scaled Magnetization_Inset_StDev Plots}. This agreement indicates that the deviation at low temperatures is not simply a result of rescaling the magnetization with new exponents or renormalizing temperature or field. Instead, the magnetization is systematically suppressed or enhanced relative to the quantum critical scaling form. Such behavior is naturally explained by the emergence of an additional energy scale and new low-energy dynamics around 3~K, consistent with a picture of spinon excitations coupled to an internal gauge field.

In summary, the magnetization of YbZn$_2$GaO$_5$ exhibits a scale invariance that resembles a zero-field QCP between 5 K and 70 K. This scale invariance breaks down at temperatures below 5 K in a well-defined temperature- and field-dependent manner. While some studies on QSL candidates suggest that a QSL phase would retain scale invariance due to the lack of any energy scale associated with long-range magnetic order, our results indicate that this breakdown of scale invariance is due to the entrance into a QSL phase with low-energy excitations and gauge field interactions that possess an inherent energy scale. Our phenomenological model supports this interpretation. Notably, spinon quasiparticles are typically only detected in spectroscopic techniques (such as INS, Raman scattering, resonant inelastic x-ray scattering) \cite{Shen2016,Feng2017,Shen2018,Luo2018,Gao2019,Li2019,Wang2019,Revelli2020,Fu2021,Ran2022,Zhao2022,Smaha2023,Jeon2024,Breidenbach2025,Han2025}, or as subtle features in torque magnetometry and heat capacity \cite{Zheng2025Thermodynamic,Zheng2025Unconventional,Shaginyan2015,Sonnenschein2019}. This study therefore offers a different, albeit indirect, method of investigating spinon excitations, their interactions with the internal gauge field, and their effects on measurable quantities including magnetization. As such, these results encourage further work on QSL candidates in applied magnetic field, including the exploration of related compounds and the behavior of their spinon excitations and gauge field interactions. This would facilitate direct comparison and improved understanding of the properties of spinon excitations. Supporting theoretical work, particularly at finite temperature and magnetic field, would greatly constrain possible QSL phases and expand our understanding of these materials to cover a wider phase diagram.

Research at Washington University was supported by the National Science Foundation (NSF) Division of Materials Research Award DMR-2236528. S.G. acknowledges the NRT LinQ, supported by the NSF under Grant No. 2152221. A portion of this work was performed at the National High Magnetic Field Laboratory (NHMFL), which is supported by National Science Foundation Cooperative Agreements No. DMR-1644779 and No. DMR-2128556 and the Department of Energy (DOE). J.S. acknowledges support from the DOE BES program "Science at 100 T". The work performed at Duke University is supported by the U.S. Department of Energy, Office of Science, Office of Basic Energy Sciences, under Award No. DE-SC0023405.

\bibliographystyle{apsrev4-2}
\bibliography{references}

\clearpage

\begin{widetext}
\begin{center}
{\large \textbf{Supplementary Information for ``Scaling Breakdown as a Signature of Spinon-Gauge Interaction in the Quantum Spin Liquid YbZn$_2$GaO$_5$''}}\\[1em]

Shannon Gould$^{1}$, John Singleton$^{2}$, Rabindranath Bag$^{3}$, Sara Haravifard$^{3,4,5}$, Sheng Ran$^{1}$ \\[0.5em]
\textit{{\small $^1$Department of Physics, Washington University in St. Louis, St. Louis, Missouri 63130, USA}\\
{\small $^2$National High Magnetic Field Laboratory, Los Alamos National Laboratory, Los Alamos, New Mexico 87545, USA}\\
{\small $^3$Department of Physics, Duke University, Durham, North Carolina 27708, USA}\\
{\small $^4$Department of Mechanical Engineering and Materials Science, Duke University, Durham, North Carolina 27708, USA}\\
{\small $^5$Department of Electrical and Computer Engineering, Duke University, Durham, North Carolina 27708, USA}} \\[0.5em]

\end{center}
\end{widetext}

\preprint{APS/123-QED}

\title{SI: Scaling Breakdown as a Signature of Spinon-Gauge Interaction in the Quantum Spin Liquid YbZn$_2$GaO$_5$}

\author{
Shannon Gould$^{1}$,
John Singleton$^{2}$,
Rabindranath Bag$^{3}$,
Sara Haravifard$^{3,4,5}$,
Sheng Ran$^{1}$ \\
\textit{{\small $^1$Department of Physics, Washington University in St. Louis, St. Louis, Missouri 63130, USA} \\
{\small $^2$National High Magnetic Field Laboratory, Los Alamos National Laboratory, Los Alamos, New Mexico 87545, USA} \\
{\small $^3$Deparment of Physics, Duke University, Durham, North Carolina 27708, USA} \\
{\small $^4$Department of Mechanical Engineering and Materials Science, Duke University, Durham, North Carolina 27708,~USA}
{\small $^5$Department of Electrical and Computer Engineering, Duke University, Durham, North Carolina 27708, USA}
}}

\date{\today}

\maketitle

\section{\label{sec:level1}Extraction Magnetometry}

We performed extraction magnetometry on YbZn$_2$GaO$_5$ in pulsed magnetic fields of up to 60~T and temperatures down to 0.6 K. Magnetization was obtained in magnetic fields oriented parallel and perpendicular to the c-axis. The extraction magnetometry probe held samples within a 1-mm-diameter cylindrical ampoule. Single crystals of YbZn$_2$GaO$_5$ were cleaved and stacked in order to fit in the ampoule in the desired orientation.

The extraction magnetometer involved a triple compensation technique to optimize the sample signal. First, an empty compensated coil with 1000 turns in one direction and 500 turns in the other direction was adjusted to produce no signal induced by the $dH/dt$ of the pulsed field. An additional single turn compensation coil is wound on the outside of the main coil, and a fraction of the voltage induced by this coil is added to or subtracted from the main coil induced voltage in order to reduce the residual voltage. Finally, each measurement involves two field pulses so that the sample signal can be subtracted from the background (no sample) signal. Setup for the measurements also included shim adjustment in order to optimize the vertical positioning of the sample inside the compensation coil. Using shims of thicknesses 0.01", 0.02", 0.03", and 0.04", the sample signal was measured in $10$ T test pulses to find the shim thickness that produced the largest sample signal.

The raw voltage signal measured from the pickup coil was converted to magnetization using previously collected data from vibrating sample magnetometry (VSM) in magnetic fields of up to $14$ T. Conversion factors were obtained from the $20$ K datasets for each field direction.

\section{van Vleck Paramagnetism Contribution}
For Yb$^{3+}$, the crystal electric field (CEF) ground state is a Kramer's doublet, so the saturation moment corresponds to $\tfrac{1}{2}g\mu_B$, with an anisotropic $g$ factor \cite{Bag2024,Zhao2025}. We fit the high-field region ($>30$ T) of the base temperature data to a linear-in-field function, $M=\chi_{vv}H+gJ_{eff}$ with $J_{eff}=1/2$. From this fit, we extracted the van Vleck susceptibilities, $\chi_{vv}$, and the anisotropic $g$ factors. We obtain $\chi_{vv}^c=0.0097$ $\mu_B/Yb/T$ and $g_c=3.60$ for $H||c$, $\chi_{vv}^{ab}=0.0033$ $\mu_B/Yb/T$ and $g_{ab}=3.20$ for $H||ab$. 
These values for the anisotropic Lande $g$ factors agree well with those obtained previously in fields of up to 14 T \cite{Bag2024}.
In our subsequent scaling analyses, Fig. 2 (main text) and Fig. \ref{fig:ScaledSusceptibility} (SI), we refer to this van Vleck contribution as $\chi_{vv}$ or $M_{vv}$, where $M_{vv}=\chi_{vv}(\mu_0 H)$. This contribution is subtracted from the measured $M$ or $\chi$ to obtain the critical contributions to the magnetization or susceptibility.

\begin{figure}
\centering
\includegraphics[width=0.48\textwidth]{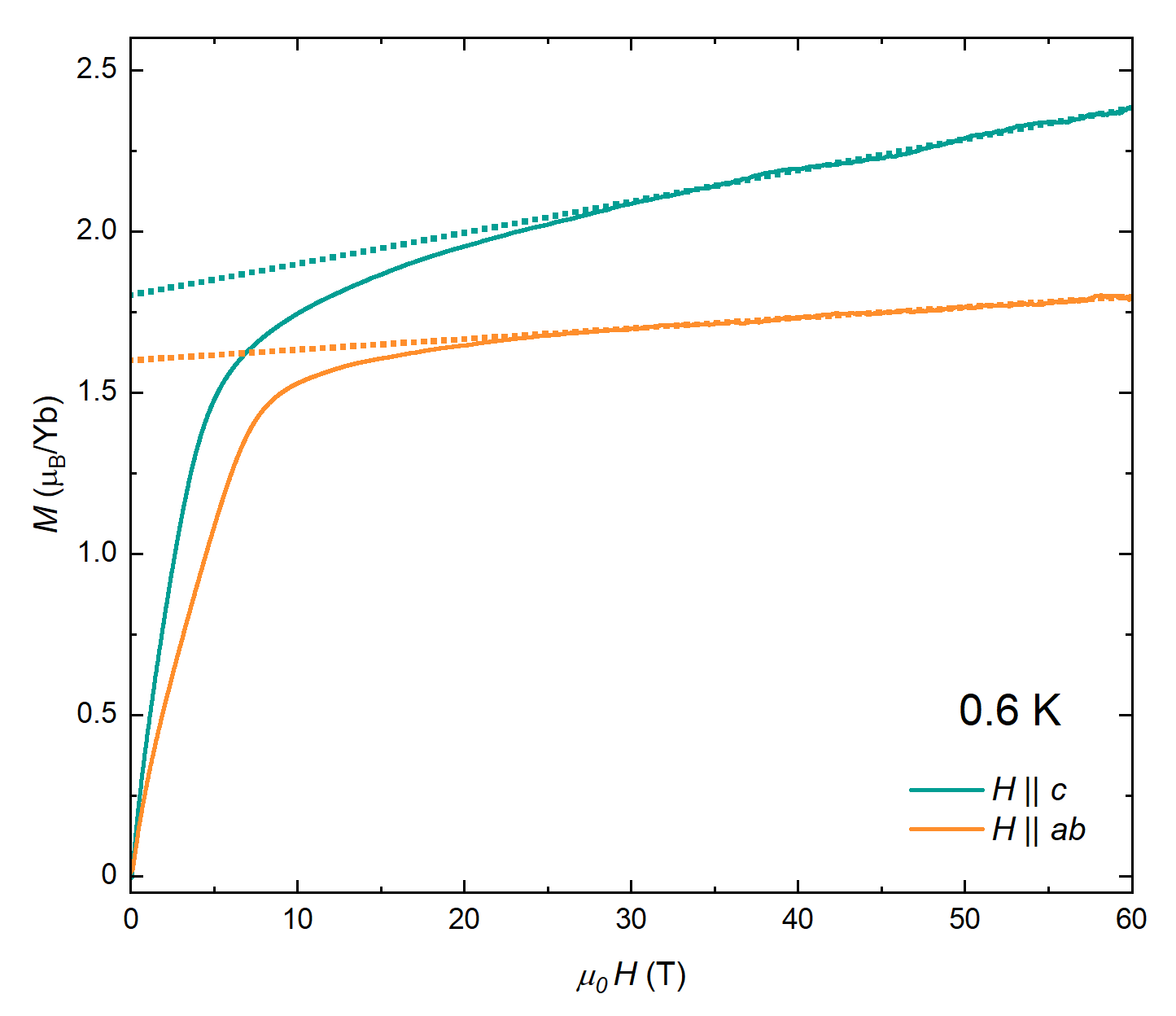}
\caption{High field $M(H)$ fit to obtain the van Vleck contribution. The fit was performed on the base temperature (0.6 K) data for both field orientations.}
\label{fig:Mvv}
\end{figure}

\section{Susceptibility}
\begin{figure}
\centering
\includegraphics[width=0.48\textwidth]{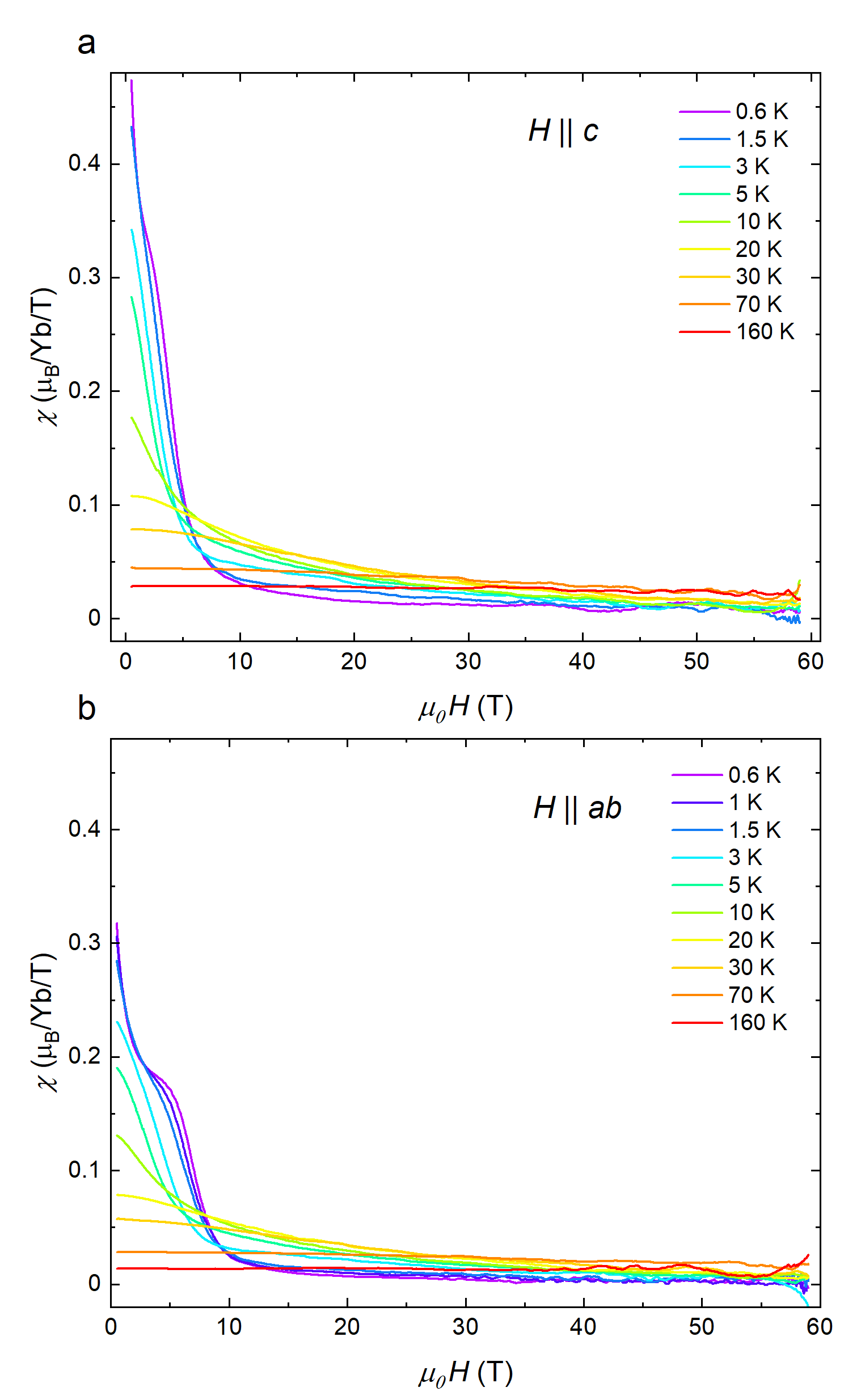}
\caption{$\chi(H)$ for $H||c$ (a) and $H||ab$ (b).}
\label{fig:XvsH}
\end{figure}

The magnetic susceptibility was calculated from the raw $M(H)$ data as $\chi=\partial M/\partial H$. $\chi(H)$ is shown in Fig. \ref{fig:XvsH} for $H||c$ (a) and $H||ab$ (b).

\section{Brillouin Function Fit}
\begin{figure}
\centering
\includegraphics[width=0.48\textwidth]{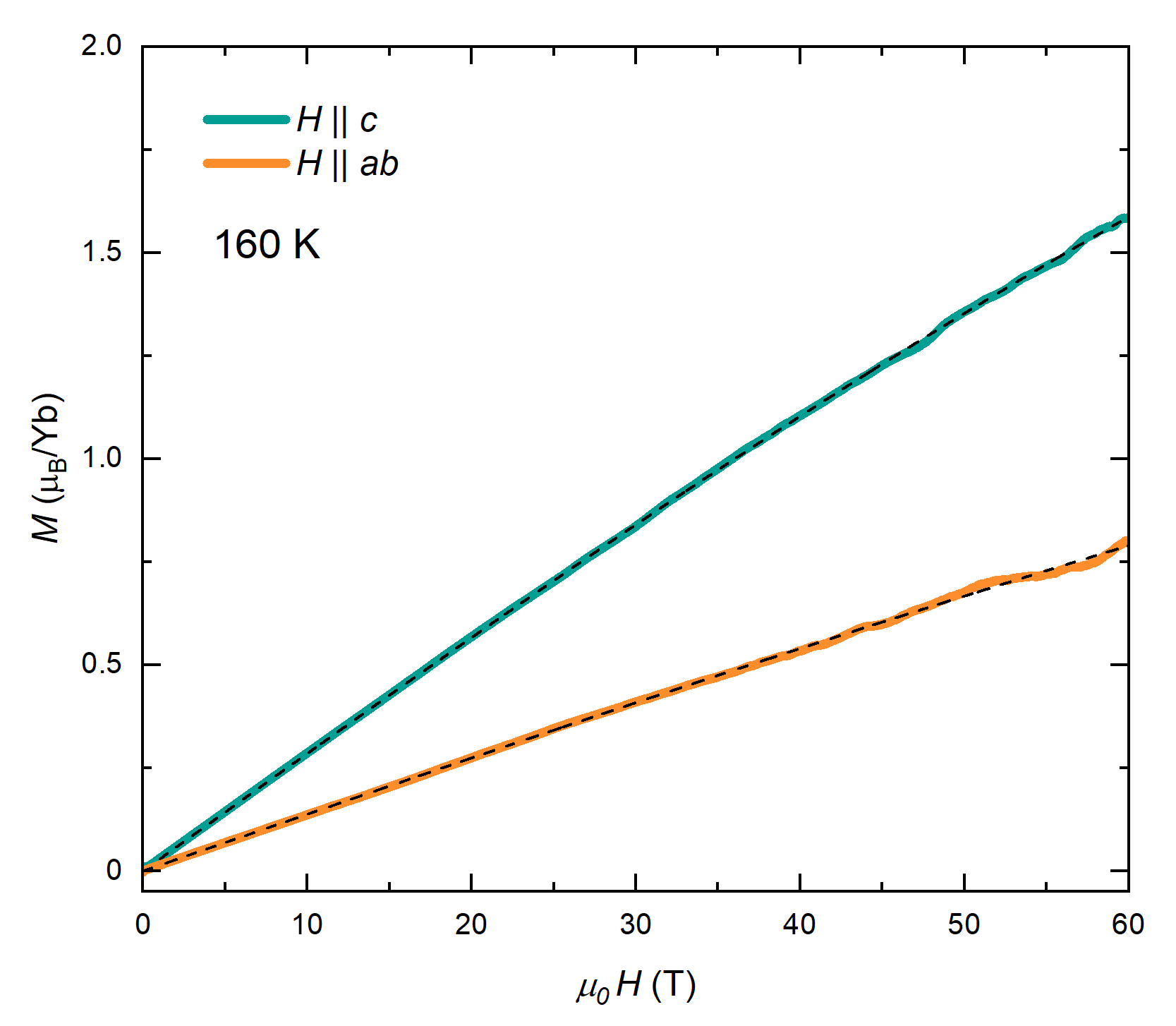}
\caption{$M(H)$ fit to the Brillouin function for $J = 1/2$ and $T = 160$ K. The dashed line shows the successful fit.}
\label{fig:Brillouin_Fit}
\end{figure}

We conducted Brillouin function fits to all $M(H)$ data, using
\[
M=Ng\mu_BJ  \operatorname{tanh} \left( \frac{gJ\mu_BB}{k_BT} \right), 
\]
with $J=1/2$ due to YbZn$_2$GaO$_5$ being an effective spin-1/2 system \cite{Bag2024}. Fits were unsuccessful for $T \leq 70$ K and both field directions. We show the successful fit for $T=160$ K in Fig. \ref{fig:Brillouin_Fit}. 

\section{Results of Scaling Analysis on $M(H)$}
In the scaling analysis shown in Fig. 2 (main text), we first assumed $\nu z = 1$ and $H_c = 0$. To refine these parameters for $H||c$ and ($H||ab$) respectively, we fixed $\gamma = 0.762$ (0.774) and $H_c=0$, and optimized $\nu z$, obtaining $\nu z = 0.98$ (0.96) for $H||c$ ($H||ab$). Further optimization of $H_c$ with $\gamma = 0.762$ (0.774) and $\nu z = 0.98$ (0.96) yielded $\mu_0 H_c=0$ T for both field orientations. In each optimization step, we calculated the mean standard deviation ($\sigma$) of curves for different values of $\gamma$, $\nu z$, $H_c$. The final, optimal value was determined to be that which minimized the standard deviation. These standard deviation plots are shown in the main text, Fig. 2c-e for $H||c$ and Fig. 2f-h for $H||ab$. These results confirm that the scaling behavior is consistent with $\nu z \approx 1$ and $\mu_0 H_c = 0$ T.

\section{Scaled Susceptibility}
\begin{figure*}
\centering
\includegraphics[width=0.98\textwidth]{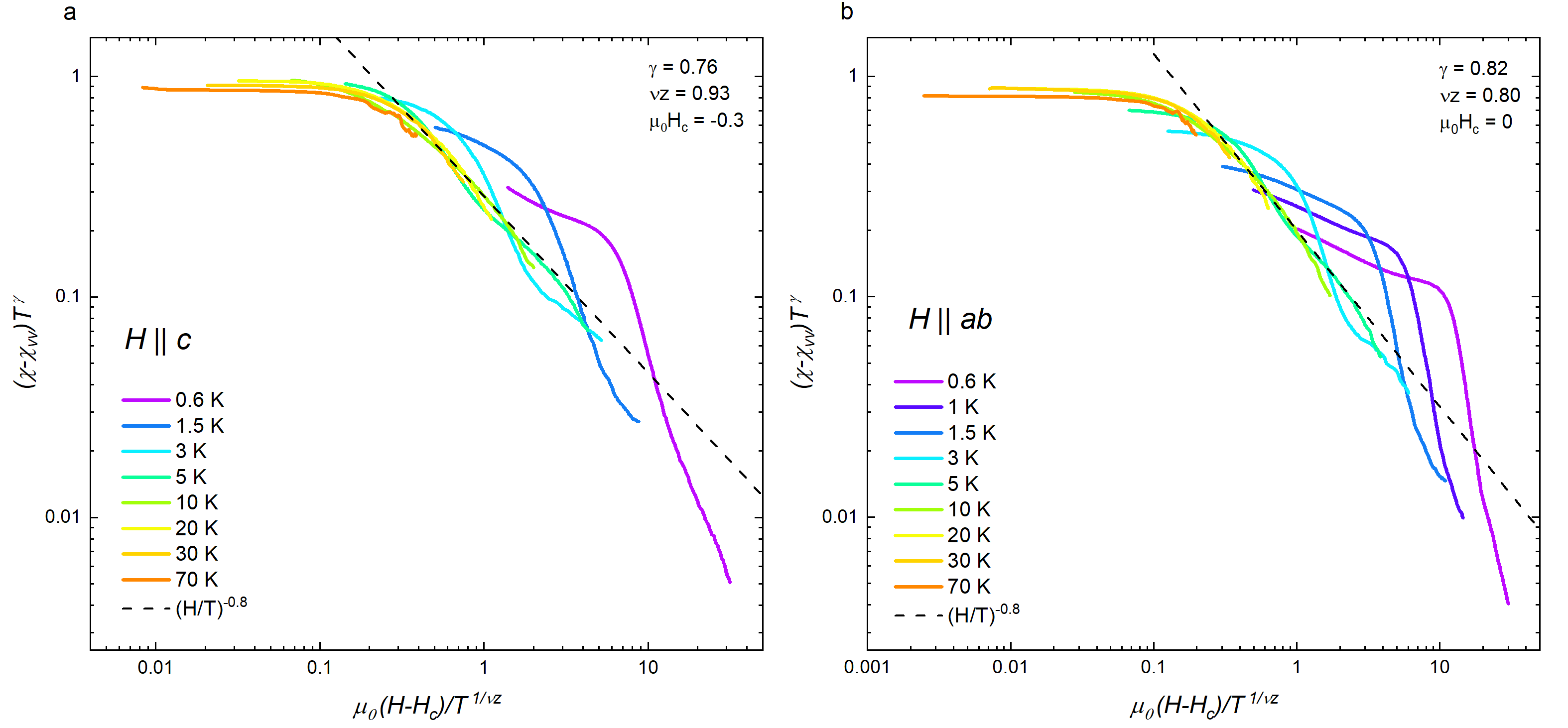}
\caption{Results of a scaling analysis on the susceptibility, $\chi=\partial M/\partial H$, for both $H||c$ (a) and $H||ab$ (b). The black dashed line indicates the $(H/T)^{-0.8}$ curve, showing good agreement with the high $H/T$ region of the $5$ K $-$ $70$ K data for both field directions.}
\label{fig:ScaledSusceptibility}
\end{figure*}

A scaling analysis was also performed directly on $\chi(H)$ (for $5$ K $-$ $70$ K) in order to confirm the scale invariance and values of $\gamma,\nu z,H_c$. Shown in Fig. \ref{fig:ScaledSusceptibility}, this analysis used the scaling function, $(\chi-\chi_{vv})T^\gamma=\mu_0(H-H_c)/T^{1/\nu z}$, which is consistent with the scaling function used for the magnetization (see main text). We indeed observe scale invariance for curves between 5 K and 70 K. The optimal values for $\gamma,\nu z, H_c$ are in close agreement with those obtained from the scaling analysis of $M(H)$. Slight differences can be attributed to poorer signal to noise in the $\chi(H)$ data. For both field directions, the high $H/T$ dependence of the universal $\chi$ curve roughly follows $(H/T)^{-0.8}$. This further confirms the values obtained for $\gamma$ and indicates a different regime from the low $H/T$ where $(\chi-\chi_{vv})T^\gamma$ is constant.

\section{Phase Diagrams}
\begin{figure}
\centering
\includegraphics[width=0.48\textwidth]{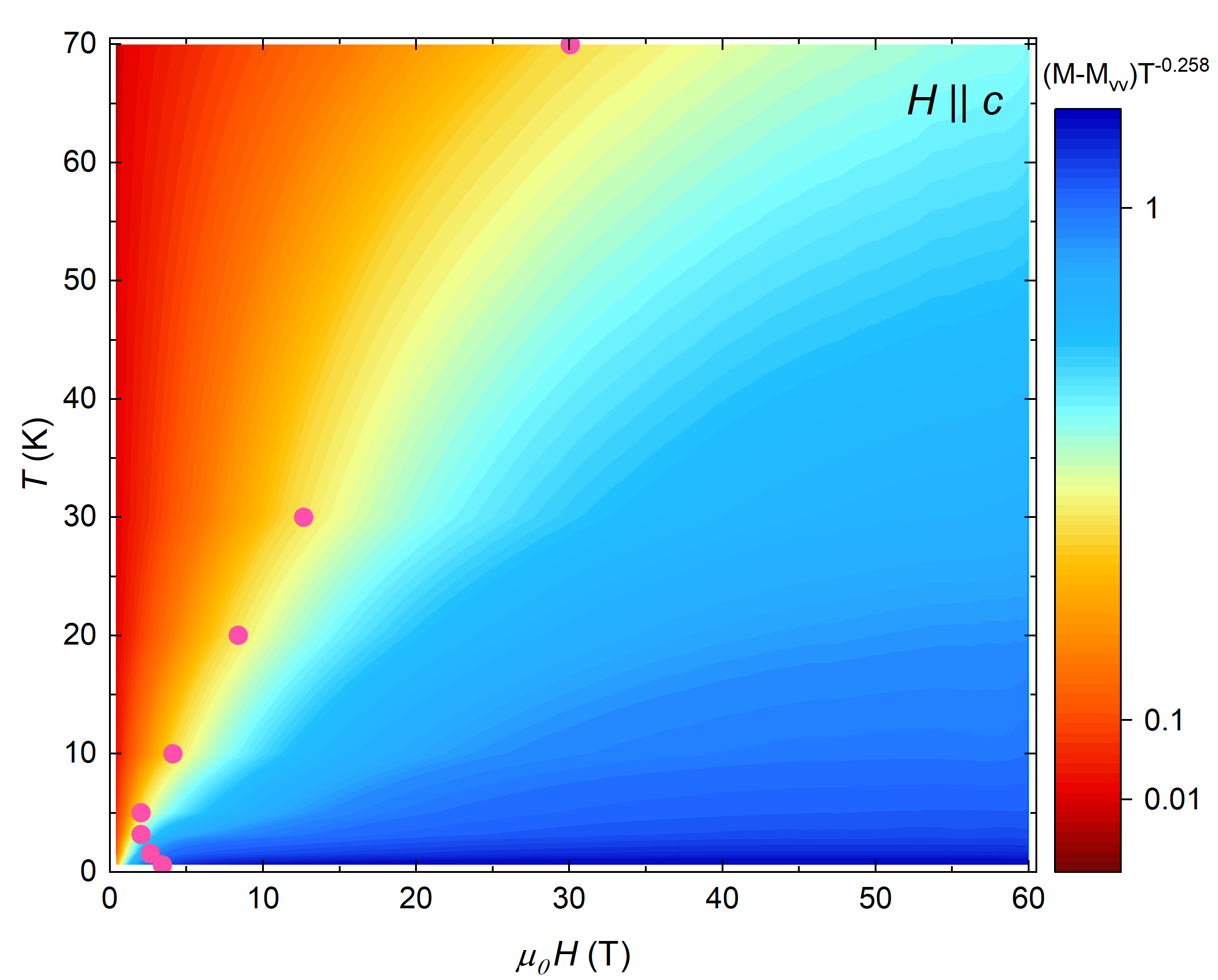}
\caption{$H-T$ phase diagram for $H||c$, over the full region studied in the experiment (excluding 160 K).}
\label{fig:HparaC Contour Full}
\end{figure}

\begin{figure}
\centering
\includegraphics[width=0.48\textwidth]{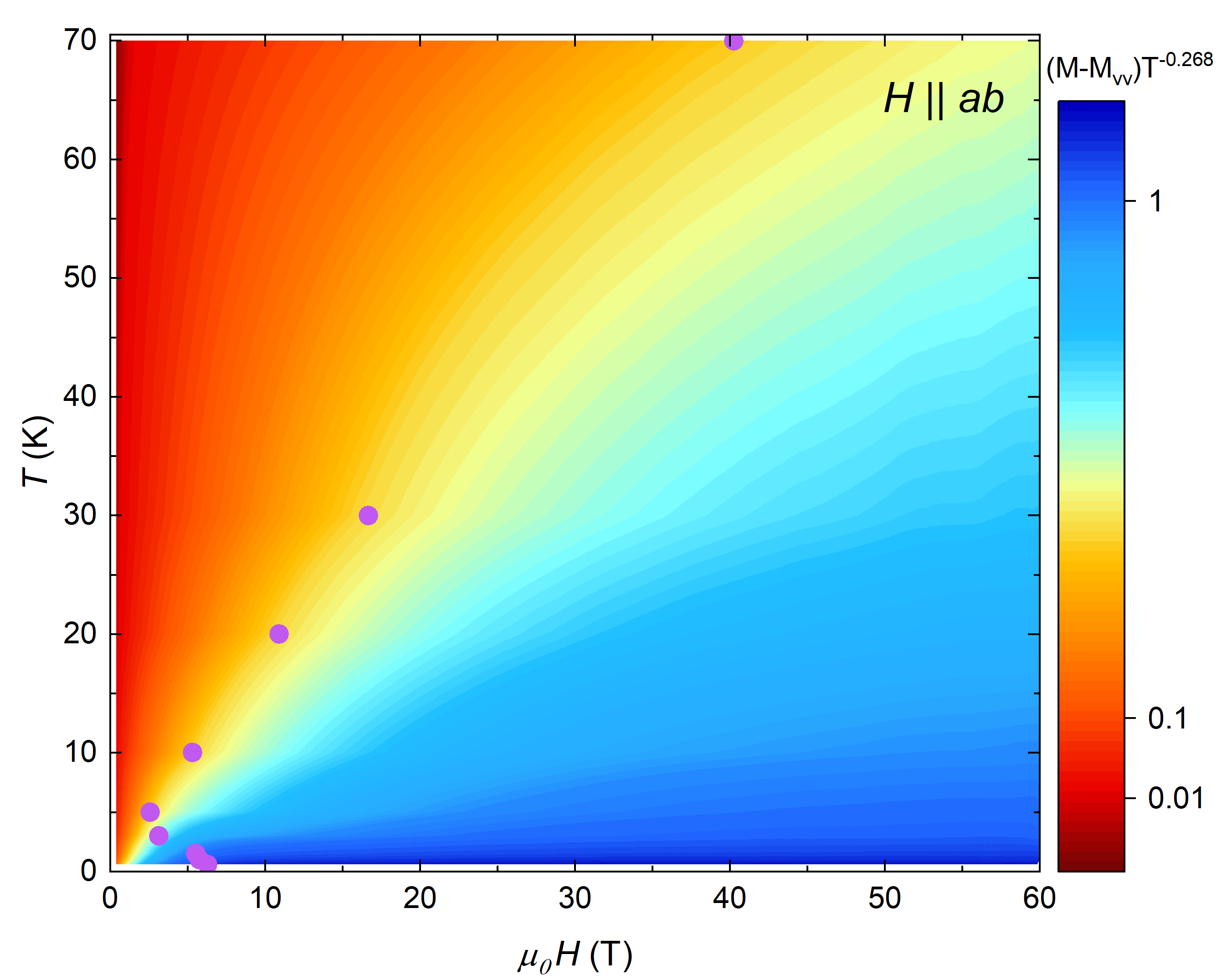}
\caption{$H-T$ phase diagram for $H||ab$, over the full region studied in the experiment (excluding 160 K).}
\label{fig:HperpC Contour Full}
\end{figure}

\begin{figure}
\centering
\includegraphics[width=0.48\textwidth]{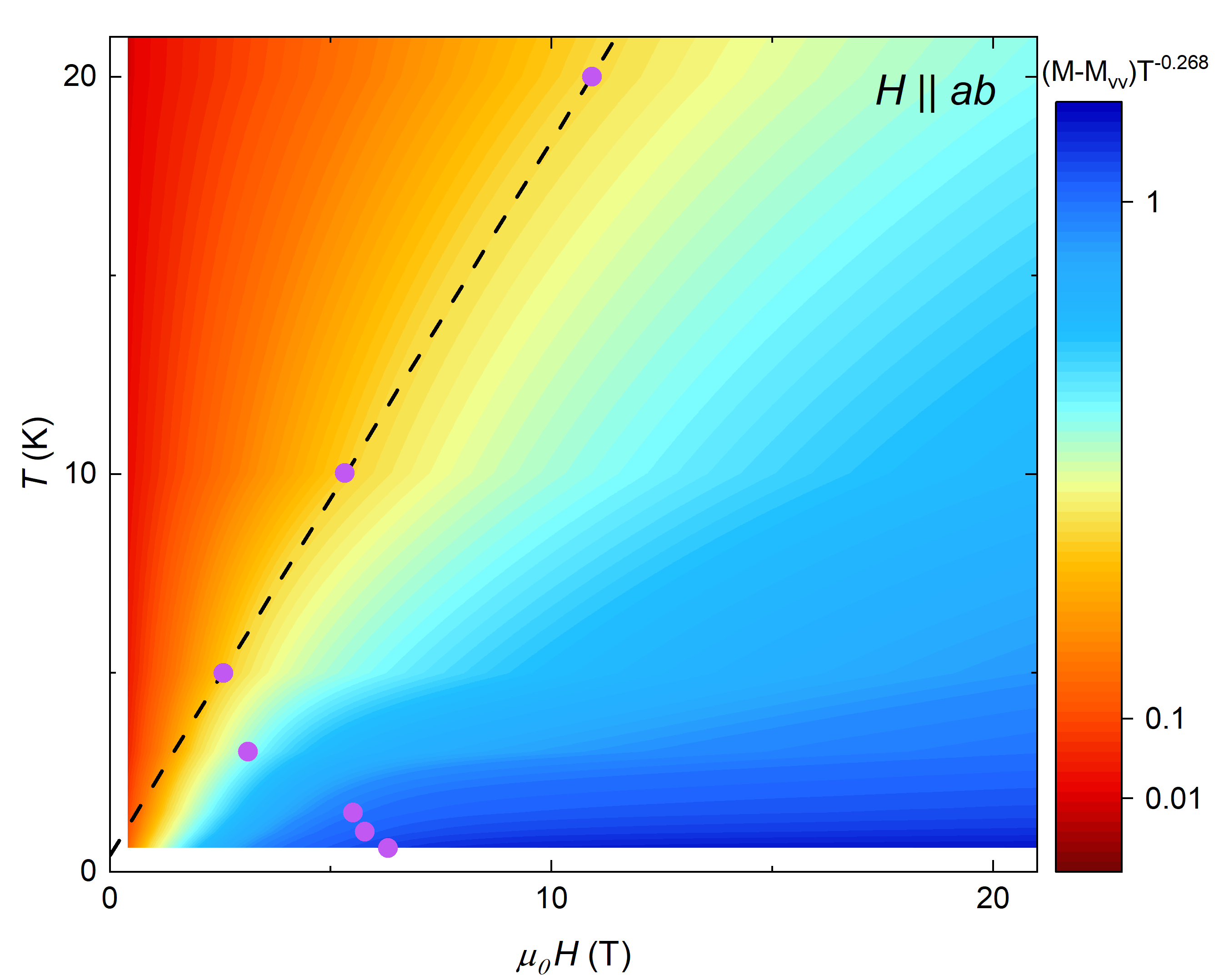}
\caption{$H-T$ phase diagram for $H||ab$, zoomed in to the low $H-T$ region. The black dashed line indicates the boundary of the QC fan, tracing back to the origin.}
\label{fig:HperpC Contour Zoom}
\end{figure}

Additional $H-T$ phase diagrams are shown in Figs. \ref{fig:HparaC Contour Full}-\ref{fig:HperpC Contour Zoom}. Fig. \ref{fig:HparaC Contour Full} shows the $H||c$ diagram over the full $H-T$ region studied in the experiment, excluding 160 K. The plot zoomed in to the low $H-T$ region is shown in the main text, Fig. 3. Figs. \ref{fig:HperpC Contour Full} and \ref{fig:HperpC Contour Zoom} show the corresponding phase diagrams for $H||ab$, spanning the full $H-T$ region and only the low $H-T$ region, respectively.

\section{Low Temperature Scaling Attempt}

\begin{figure}
\centering
\includegraphics[width=0.48\textwidth]{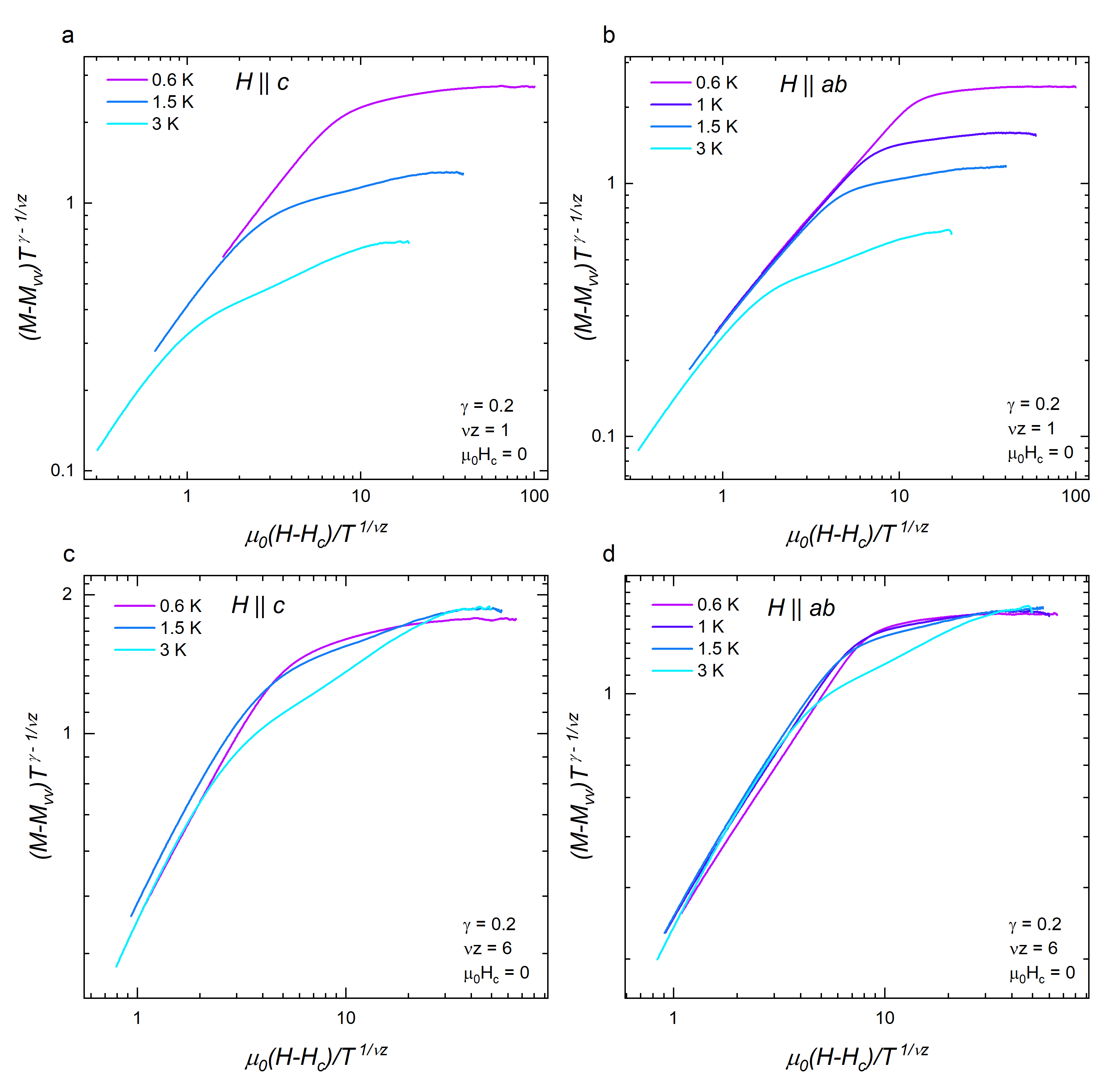}
\caption{Independent attempts to scale magnetization at temperatures $< 5$ K for $H||c$ (a, c) and for $ H||ab$ (b, d).}
\label{fig:LowT Scale Attempt}
\end{figure}

Considering the possibility of a new fixed point or change in universality class of the QCP below 3 K \cite{Vojta2003}, we attempted to scale the low temperature curves independently from the rest of the data. We attempted separate scaling analyses on the $M(H)$ data below 5 K. Fig. \ref{fig:LowT Scale Attempt}a, c show two independent attempts to scale the $H||c$ results. For different values of $\gamma,\nu z,H_c$, the data fails to collapse. It is apparent from visual inspection that the curvature and asymptotic slopes of the data are incompatible with a total collapse for any value of $\gamma,\nu z,H_c$. We observe similar behavior for $H||ab$ as shown in Fig. \ref{fig:LowT Scale Attempt}b, d.

\section{Phenomenological Model}
To describe the breakdown of scale invariance in a more quantitative manner, we introduce a phenomenological model for $M$ based on this intuitive physics picture.  In the quantum critical regime, $M$ only has the critical contribution: 
\[
M=M_{QC}=T^{-\gamma+1/\nu z}\Phi[\mu_0(H-H_c)/T^{1/\nu z}],
\]
where $\Phi(x)$ is an arbitrary scaling function. This model inherently captures the lack of any relevant energy scale in the system. We choose the scaling function to be $\Phi(x)=x _2F_1(1/2,1/2,3/2,-x^2) $ where $_2F_1(a,b,c,-x^2)$ is the hypergeometric function. This choice is based on the general shape of the universal curve in Fig. 2a-b (main text): a moderate linear slope at small-x gradually turns over to a smaller slope at large-x. We plot $MT^{\gamma-1/\nu z}$ versus $\mu_0(H-H_c)/T^{1/\nu z}$ in Fig. 4d (main text). Representative values for the critical exponents, $\gamma=0.8$, $\nu z=1, H_c=0$, are used. As expected, this model results in complete collapse of all temperatures over the full x-range.

Next, we incorporate an emergent energy scale, $\Delta$, into the model that competes with the scale invariance arising from quantum critical fluctuations. $M$ is modeled as 
\[
M=M_{QC}[1+B(T)f(X)].
\]
Two new functions, $B(T)$ and $f(X)$, encode the contributions from low-energy excitations associated with the spin liquid state. $B(T)$ controls the temperature dependence of this additional contribution. It satisfies the limiting behaviors $B(T) \rightarrow 0$ for $T \gg \Delta$ and $B(T) \rightarrow 1$ for $T \ll \Delta$, where $\Delta$ is the emergent energy scale in Kelvin. The function we have chosen to realize these limits is 
\[
B(T) = \frac{1}{1+b(T/\Delta)^p}.
\]
$b,p$ are left as adjustable parameters. The function $f(X)$ captures the field dependence of the spinon contribution, which is negative at low magnetic field and positive at high magnetic field. The function that captures this in our model is 
\[
f(X)=\frac{aX_{eff}^m}{(1+X_{eff}^m)(1+X_{eff}^d)} - c
\]
where $X_{eff}=X/X_0(T)$ 
and \[
X_0(T)=
\begin{cases}
X_0(1+d(\Delta/T)^q) & \text{if } T < \Delta \\
X_0 & \text{if } T \geq \Delta
\end{cases}
\]

Parameters available to adjust in this function include $a,c,d,m,q,X_0$. The functional forms of $B(T)$ and $f(X)$ allow some flexibility in selection of the parameters so that the details of the low temperature curves can be adjusted. However, the key features of the functions, including their limits, are fixed regardless of specific parameter values.

\end{document}